\documentclass[aps,preprint]{revtex4-1}
\usepackage{graphicx}
\usepackage{amsmath}
\usepackage{makeidx}
\usepackage{amsfonts}
\usepackage{amssymb}

\begin{document}

\title{Enhancing the efficiency of quantum annealing via reinforcement: A path-integral Monte Carlo simulation of the quantum reinforcement algorithm}

\author{A. Ramezanpour}
\email{aramezanpour@gmail.com}
\affiliation{Physics Department, College of Sciences, Shiraz University, Shiraz 71454, Iran}
\affiliation{Leiden Academic Centre for Drug Research, Faculty of Mathematics and Natural Sciences, Leiden University, Leiden, The Netherlands}
\date{\today}

\date{\today}

\begin{abstract}
The standard quantum annealing algorithm tries to approach the ground state of a classical system by slowly decreasing the hopping rates of a quantum random walk in the configuration space of the problem, where the on-site energies are provided by the classical energy function. In a quantum reinforcement algorithm, the annealing works instead by increasing gradually the strength of the on-site energies according to the probability of finding the walker on each site of the configuration space. Here, by using the path-integral Monte Carlo simulations of the quantum algorithms, we show that annealing via reinforcement can significantly enhance the success probability of the quantum walker. More precisely, we implement a local version of the quantum reinforcement algorithm, where the system wave function is replaced by an approximate wave function using the local expectation values of the system. We use this algorithm to find solutions to a prototypical constraint satisfaction problem (XORSAT) close to the satisfiability to unsatisfiability phase transition. The study is limited to small problem sizes (a few hundreds of variables), nevertheless, the numerical results suggest that quantum reinforcement may provide a useful strategy to deal with other computationally hard problems and larger problem sizes even as a classical optimization algorithm.                   
\end{abstract}


\maketitle

\section{Introduction}\label{S0}
Finding a solution to a computationally hard constraint satisfaction problem becomes more difficult for a typical instance of the problem as one approaches the phase transition from a satisfiable (SAT) to unsatisfiable (UNSAT) phase \cite{sat-nature-1999,sat-science-2002}; in the SAT phase, with high probability there is a solution to the problem satisfying all the constraints, whereas in the UNSAT phase there is no solution to the problem with high probability. A reinforcement algorithm tries to find a solution to the problem by utilizing the information that is obtained from the system at each step of the algorithm. This provides a class of powerful classical reinforcement algorithms to deal with such problems \cite{SB-book-1998,BZ-prl-2006}. In this paper, we show that adding reinforcement to the standard quantum annealing algorithm is helpful in the study of a prototypical constraint satisfaction problem. More precisely, we observe that a path-integral Monte Carlo simulation of the quantum reinforcement algorithm gives much higher success probabilities than the simulated quantum annealing algorithm.

The presence of strong and long-range correlations between the problem variables, due to a spin glass or a freezing phase transition \cite{gibbs-pnas-2007,semerjian-jstat-2008,gibbs-pre-2008,clustering-prl-2005,clustering-jstat-2008}, is responsible for the computational complexity of a constraint satisfaction problem close to the SAT-UNSAT transition. It means that to obtain efficient approximation algorithms, we should be able to extract efficiently the global information that is relevant to the problem, from the system of interacting variables. For instance, the Gaussian elimination algorithm provides an efficient way of solving a set of linear equations over binary variables, which is known as the XOR-satisfiability (XORSAT) problem \cite{xor-prl-2001,xor-pre-2001}. See also Ref. \cite{qxor-prl-2009} for a quantum algorithm for the XORSAT problem. Nevertheless, it is very difficult to write the Gaussian elimination algorithm in a form that is amenable to local message-passing algorithms \cite{MM-book-2009}. Another example in this direction is provided by Ref. \cite{entropy-jstat-2016}, where the entropy, or number of solutions in a region around a point in the configuration space, is estimated at each step to guide the search algorithm. Here, the entropy is playing the role of the global information that is used by the algorithm. The main problem is that obtaining good estimations of the relevant global quantities and writing this computation in a locally manageable way is usually difficult. There are, however, special examples, where the global constraints can be treated exactly and efficiently via message passing along a spanning tree of the interaction graph \cite{globalgame-2011,sign-prb-2012}.

In a previous study \cite{QR-pra-2017}, we introduced a quantum reinforcement algorithm, which uses the global information contained in the wave function of the system in a quantum annealing algorithm. More precisely, we considered a continuous-time quantum random walk in the configuration space of the classical optimization problem \cite{ALZ-pra-1993,K-cp-2003,A-jqi-2003}. At the beginning of the algorithm, the on-site energies at each point of the configuration space are given by the energy function of the classical problem. These on-site energies are gradually modified according to the wave function of the evolving quantum system to localize preferentially the wave function on a solution to the classical problem. Using exact numerical simulations of small systems, we showed that such quantum feedback increases the minimal energy gap of the quantum system in a quantum annealing algorithm, and therefore could be useful in the study of hard optimization problems \cite{F-sci-2001,NC-book-2002}.

Notice that the quantum reinforcement algorithm results in a nonlinear Schrodinger equation, and it is known that one can efficiently solve a computationally hard problem with nonlinear quantum mechanics \cite{lloyd-prl-1998}. In addition, we know that the standard quantum annealing algorithm is frustrated by the exponentially small energy gaps of the system in the annealing process \cite{AHJ-pnas-2010,qxor-prl-2010,qxor-pre-2011,qxor-pra-2012}. There are remedies to this problem that work by adding auxiliary interactions to the Hamiltonian to suppress the spoiling quantum transitions in the annealing process \cite{B-jpa-2009,C-prl-2013}. These auxiliary interactions are highly nonlocal, but good approximation algorithms can still be obtained by replacing the nonlocal Hamiltonians with effective local Hamiltonians \cite{localCA-pra-2014,localCA-pnas-2017}.

In this paper, we show that the local versions of the quantum reinforcement algorithm work also for larger problem sizes. To this end, we resort to quantum Monte Carlo simulations of the algorithm, using the path-integral representation of the quantum system at equilibrium for sufficiently low temperatures \cite{tosatti-prb-2002,pathMC-prb-2008,QC-prep-2013}. We apply the algorithm to the XORSAT problem close to the SAT-UNSAT phase transition, where the problem is expected to be hard for a local algorithm. We compare the performance of the quantum reinforcement algorithm with that of the standard quantum annealing algorithm for problems with a few hundreds of variables. We observe considerable improvements in the success probability of the algorithms by adding reinforcement to the quantum annealing algorithm. Note that our previous study \cite{QR-pra-2017} was limited to small problem sizes and exact numerical simulations of a fully connected spin-glass model. 
Moreover, in that study we could not observe the superior performance of the quantum reinforcement algorithm in larger systems, compared to the standard quantum annealing.

The paper is organized as follows. In Sec. \ref{S1} we define the problem in more detail. Then we briefly review the quantum reinforcement algorithm and its local approximations in Sec. \ref{S2}. The path-integral Monte Carlo simulation of the algorithms is described in Sec. \ref{S3}. Section \ref{S4} is devoted to the presentation of the numerical results, and finally Sec. \ref{S5} gives the conclusions.

\section{Problem statement and definitions}\label{S1}
We consider the classical optimization problem of minimizing an energy function $E(\boldsymbol\sigma)$ of $N$ binary spins $\sigma_i=\pm 1$. As the benchmark, we take the random regular XORSAT problem \cite{MM-book-2009}, with
\begin{align}
E(\boldsymbol\sigma)=\sum_{a=1}^M(1-J_a\prod_{i\in \partial a}\sigma_i). 
\end{align}
Here, $M$ is the number of $K$-spin interactions and $J_a=\pm 1$ with equal probability. The subset of spins involved in interaction $a$ are denoted by $\partial a$. The $M$ interactions are selected randomly and uniformly from the set of all possible $K$-spin interactions. The interaction graph is regular in the sense that each interaction term involves exactly $K$ spins, and each spin is associated with exactly $L$ interactions.

A solution to this problem is a spin configuration with energy zero, where $J_a\prod_{i\in \partial a}\sigma_i=1$ for all the $a$. The problem is called satisfiable if there is at least one solution to the problem. It is well known that the problem is satisfiable (SAT) with high probability for $L<K$, and unsatisfiable (UNSAT) for $L>K$ \cite{MM-book-2009}. Moreover, the problem is computationally easy and belongs to the complexity class $P$; this means that we can decide if the problem is SAT or UNSAT in a computation time that grows polynomially with the size of the problem ($N$). In addition, as long as the problem is satisfiable, a solution can easily be obtained by the Gaussian elimination algorithm.  To be specific, we consider random regular XORSAT problems with parameters $(K=4,L=3)$. We know that for these values of $K$ and $L$ the solution space is clustered and it is computationally difficult to find a solution by a local algorithm such as the Markov Chain Monte Carlo \cite{xor-prl-2001,xor-pre-2001,MM-book-2009}. It is also known that we need an exponentially large computation time to find the ground state of the XORSAT problem by the standard quantum annealing algorithm \cite{qxor-prl-2010,qxor-pre-2011,qxor-pra-2012}.

We shall use a continuous-time quantum random walk to explore the space of spin configurations $\boldsymbol\sigma=\{\sigma_1,\dots,\sigma_N\}$. The space is a hypercube of $2^N$ sites corresponding to the total number of spin configurations. The Hamiltonian for a particle walking in the energy landscape of the classical optimization problem is given by
\begin{align}
H=\sum_{\boldsymbol\sigma} E(\boldsymbol\sigma) |\boldsymbol\sigma\rangle\langle \boldsymbol\sigma|-\sum_{\boldsymbol\sigma}\sum_{i=1}^N\Gamma\left( |\boldsymbol\sigma^{-i}\rangle\langle \boldsymbol\sigma| + |\boldsymbol\sigma\rangle\langle \boldsymbol\sigma^{-i}| \right). 
\end{align}
The parameter $\Gamma$ determines the strength of tunneling from $|\boldsymbol\sigma \rangle$ to a neighboring state $|\boldsymbol\sigma^{-i} \rangle$. Here, $|\boldsymbol\sigma^{-i} \rangle$ denotes the spin state which is different from $|\boldsymbol\sigma \rangle$ only at site $i$. In terms of the quantum spin variables (Pauli matrices), the above Hamiltonian reads as follows,
\begin{align}
H=\sum_a(1-J_a\prod_{i\in \partial a}\sigma_i^z)-\sum_i \Gamma \sigma_i^x. 
\end{align}
The basis states $|\boldsymbol\sigma\rangle$ are the $N$-spin states with definite $\sigma_i^z$ values, that is, $\sigma_i^z|\boldsymbol\sigma\rangle=\sigma_i|\boldsymbol\sigma\rangle $.  

Starting from an initial state $|\psi(0)\rangle$, the time evolution of the isolated system is governed by the Schrodinger equation $\hat{i}\frac{d}{dt} |\psi(t)\rangle = H |\psi(t)\rangle$ with $\hbar=1$. In the following, we shall assume that the system is always in thermal equilibrium with a thermal bath at a sufficiently small temperature. At equilibrium, the physical properties of the system are obtained from the quantum partition function $Z=\mathrm{Tr} e^{-\beta H}$, for a large inverse temperature $\beta$.

\section{Quantum Reinforcement Algorithm}\label{S2}
In this section we briefly review the quantum reinforcement algorithm introduced in Ref. \cite{QR-pra-2017}. The goal is to find a solution to the classical optimization problem by following the time evolution of the quantum system. A quantum annealing (QA) algorithm \cite{F-sci-2001} starts from the ground state of $H_x\equiv -\sum_i \Gamma \sigma_i^x$ and changes slowly the Hamiltonian to $H_c \equiv \sum_a(1-J_a\prod_{i\in \partial a}\sigma_i^z)$. The adiabatic theorem then ensures that in the absence of level crossing, the system follows the instantaneous ground state of the time dependent Hamiltonian $H_{QA}(t)= s(t)H_c+ [1-s(t)]H_x$.  
The annealing parameter $s(t)$ changes slowly from zero at $t=0$ to one at $t=t_{max}$. In the following, we shall assume that $s(t)=t/t_{max}$.

In a quantum reinforcement (QR) algorithm, we add a reinforcement term to the Hamiltonian which favors the spin states of higher probability \cite{QR-pra-2017}. More precisely, the Hamiltonian is $H_{QR}(t)= H_c+ H_x+H_r(t)$, where the reinforcement term reads as follows,  
\begin{align}
H_r(t)\equiv -r(t)\sum_{\boldsymbol\sigma}|\psi(\boldsymbol\sigma;t)|^2 |\boldsymbol\sigma\rangle\langle \boldsymbol\sigma|.
\end{align}
Here, $\psi(\boldsymbol\sigma;t)$ refers to the wave function of the quantum system.
The reinforcement parameter $r(t)$ is zero at the beginning and is expected to grow slowly with time.

\subsection{Local approximations of the algorithm}\label{S21}
To obtain a local version of the QR algorithm, we first replace the $|\psi(\boldsymbol\sigma;t)|^2$ with $\log |\psi(\boldsymbol\sigma;t)|^2$, which is an increasing function of the probability distribution. On the other hand, we can always write $\psi(\boldsymbol\sigma;t)=\exp(\sum_i K_i\sigma_i/2+\sum_{i<j}K_{ij}\sigma_i\sigma_j/2+\cdots)/\sqrt{Z}$, taking into account all the possible multispin interactions with complex couplings $K_i=K_i^R+\hat{i}K_i^I, K_{ij}=K_{ij}^R+\hat{i}K_{ij}^I, \dots$.  Consequently, $|\psi(\boldsymbol\sigma;t)|^2=\exp(\sum_i K_i^R\sigma_i+\sum_{i<j}K_{ij}^R\sigma_i\sigma_j+\cdots)/Z$ and $Z$ is the normalization constant. The coupling parameters $K_i^R, K_{ij}^R, \dots$ can in principle be determined from the expectation values $\langle \sigma_i^z\rangle, \langle \sigma_i^z\sigma_j^z\rangle, \dots$ \cite{inverse-advanc-2017}. A one-local quantum reinforcement ($1$-lQR) algorithm then is obtained by approximating the wave function with a product state,  
\begin{align}
H_r^{local}(t) \equiv -r(t) \sum_{\boldsymbol\sigma}\sum_i K_i^R\sigma_i|\boldsymbol\sigma\rangle\langle \boldsymbol\sigma|.
\end{align}
The reinforcement fields $K_i^R$ depend on the average spin values $m_i^z=\sum_{\boldsymbol\sigma} \sigma_i |\psi(\boldsymbol\sigma;t)|^2$ through $K_i^R=\frac{1}{2}\log((1+m_i^z)/(1-m_i^z))$.
More accurate approximations of the wave function and the quantum reinforcement algorithm can be obtained by considering the two-spin interactions in the expansion. This gives a two-local quantum reinforcement ($2$-lQR) algorithm. Similarly, one obtains the higher-order approximations.  The interaction pattern of the random regular XORSAT problem, however, suggests a $K$-local reinforced Hamiltonian, where     
\begin{align}
H_r^{local}(t) \equiv -r(t) \sum_{\boldsymbol\sigma}\left(\sum_i K_i^R\sigma_i+\sum_a K_a^R\prod_{i\in \partial a}\sigma_i \right)|\boldsymbol\sigma\rangle\langle \boldsymbol\sigma|.
\end{align}
In the following, we shall focus mainly on the $1$-lQR algorithm.

\section{The simulated quantum reinforcement algorithm}\label{S3}
Let us consider the one-local QR Hamiltonian $H_{QR}(t)= H_c+ H_x+H_r^{local}(t)$ with $H_r^{local}(t) = -r(t)\sum_i K_i^R \sigma_i^z$. In the following, we ignore the constant term in the energy function of the classical problem. Using the Suzuki-Trotter decomposition for the partition function $Z_{QR}=\mathrm{Tr} \exp(-\beta H_{QR})$, we get
\begin{multline}
Z_{QR}=\sum_{\vec\sigma_1,\dots,\vec\sigma_N}\exp \left(-\frac{\beta}{N_s}\sum_{\alpha=1}^{N_s}[E(\boldsymbol\sigma(\alpha))-r(t)\sum_{i=1}^NK_i^R\sigma_{i}(\alpha)] \right) \times \\
\prod_{\alpha=1}^{N_s}\langle \boldsymbol\sigma(\alpha) |e^{\frac{\beta}{N_s}\Gamma\sum_{i=1}^N\sigma_{i}^x(\alpha)}|\boldsymbol\sigma(\alpha+1)\rangle.
\end{multline}
Here, $\alpha=1,\dots,N_s$ shows different imaginary times, and $N_s$ is the number of imaginary-time slices. Note that  we are using the periodic boundary condition, i.e., $\boldsymbol\sigma(N_s+1)=\boldsymbol\sigma(1)$. The bold symbols $\boldsymbol\sigma(\alpha)$ show the spin values $\sigma_i(\alpha)$ for a given imaginary time $\alpha$.  On the other hand, the vector $\vec\sigma_i$ displays the spin values at site $i$ for different imaginary times.

Specifically, the partition function for our problem can be written as
\begin{multline}
Z_{QR}=\sum_{\vec\sigma_1,\dots,\vec\sigma_N}\exp \left(\tau\sum_{\alpha=1}^{N_s}[\sum_aJ_a\prod_{i\in \partial a}\sigma_{i}(\alpha)+r(t)\sum_iK_i^R\sigma_{i}(\alpha)] \right) \times \\
\prod_i\prod_{\alpha}\left(\cosh(\tau\Gamma)\delta_{\sigma_i(\alpha+1),\sigma_i(\alpha)}+\sinh(\tau\Gamma)\delta_{\sigma_i(\alpha+1),-\sigma_i(\alpha)}\right),
\end{multline}
where we defined $\tau\equiv \beta/N_s$. This defines a positive probability measure for the spin configuration (for positive $\Gamma$) which can be used in a standard Monte Carlo (MC) simulation. In each step of the Monte Carlo, we replace the imaginary spin values $\vec\sigma_i$ with $\vec\sigma_i'$, which is sampled from the following probability distribution,
\begin{multline}
P_{QR}(\vec\sigma_i) \propto \exp\left(\tau\sum_{\alpha=1}^{N_s}[\sum_{a\in \partial i}J_a\prod_{j\in \partial a}\sigma_j(\alpha)+r(t)K_i^R\sigma_i(\alpha)]\right)\\ \times \prod_{\alpha}\left(\cosh(\tau\Gamma)\delta_{\sigma_i(\alpha+1),\sigma_i(\alpha)}+\sinh(\tau\Gamma)\delta_{\sigma_i(\alpha+1),-\sigma_i(\alpha)}\right).
\end{multline}
This is a one-dimensional problem and the new configuration can easily be obtained by the transfer-matrix method \cite{QC-prep-2013}. Here, we use the belief propagation (BP) algorithm for this task. The BP algorithm is explained with more details in the Appendix. More precisely, the new spin values $\vec\sigma_i'$ are obtained one by one with a decimation algorithm; at each step the value $\sigma_i'(\alpha)$ is sampled from the marginal probability distribution $\mu_{\alpha}(\sigma)$, which is computed by the BP algorithm conditioned on the values of the previously decimated spins. In each Monte Carlo sweep, the $N$ spin vectors $\vec\sigma_i$ are chosen in a random sequential way and are updated according to the above procedure.

Having a quantum Monte Carlo simulation, the simulated QR algorithm starts with a random spin configuration $\{\vec\sigma_1,\dots,\vec\sigma_N\}$, where $\sigma_i(\alpha)=\pm 1$ with equal probability. We set the reinforcement parameter $r(t)=0$ and couplings $K_i^R(t)=0$, at time step $t=0$. Then, for each time step $t=1,\dots,t_{max}$ we do the following: 

\begin{enumerate}

\item Perform $t_{eq}$ Monte Carlo sweeps for equilibration. 

\item Use the last $t_{av}$ sweeps to estimate the averages $m_i=\sum_{\alpha}\sigma_i(\alpha)/N_s$.     

\item Update the reinforcement couplings $K_i^R(t)=\frac{1}{2}\log((1+m_i)/(1-m_i))$. 

\item Increase the reinforcement parameter $r(t)=r(t-1)+\delta r$.

\item Compute $E(\boldsymbol\sigma(\alpha))$ for $\alpha=1,\dots,N_s$. 

\item Report the minimum energy $E_{min}(t)=\min_{\alpha}E(\boldsymbol\sigma(\alpha))$ and stop if $E_{min}(t)=0$.

\end{enumerate}

The partition function for the $K$-local QR Hamiltonian is obtained simply by adding the extra reinforcement term, i.e., $-r(t)\sum_a K_a^R\prod_{i\in \partial a}\sigma_i(\alpha)$, to the energy function of the replicated system $E(\boldsymbol\sigma(\alpha))$. 
The simulation of the $K$-local QR algorithm is similar to the $1$-local QR algorithm except in steps $2$ and $3$. Here, in addition to the $m_i$ in step $2$, we need also to compute the average values $m_a=\sum_{\alpha}\prod_{i\in \partial a}\sigma_i(\alpha)/N_s$, and in step $3$, we have to solve the inverse problem of computing the $K_i^R$ and $K_a^R$ from the expectation values $m_i$ and $m_a$. In the Appendix, we describe an approximate algorithm to deal with this inverse problem \cite{inverse-advanc-2017}. The idea is to start from $K_i^R(old)=K_a^R(old)=0$ and change slightly the parameters depending on the difference in the associated expectation values, i.e. $K_{i,a}^R(new)=K_{i,a}^R(old)+\eta(m_{i,a}-m_{i,a}(old))$, for a positive and small $\eta$. We compute the expectation values $m_{i,a}(old)$ by the BP algorithm with the parameters $K_{i,a}^R(old)$. After each step the old parameters are replaced with the new ones, and the process is repeated for $t_{inv}$ steps.

For comparison, we also simulate the standard quantum annealing algorithm with Hamiltonian $H_{QA}(t)= s(t)H_c+ [1-s(t)]H_x$. Here, we do not have the reinforcement terms in the energy function of the replicated system. Instead the energy function and $\Gamma$ are replaced with $s(t)E(\boldsymbol\sigma(\alpha))$ and $[1-s(t)]\Gamma$, respectively. The algorithm is similar to but simpler than the $1$-local QR algorithm, in that steps $2-4$ are replaced with one step which updates $s(t)=t/t_{max}$. As before, we start from a random spin configuration. Note that at the beginning of the algorithm ($t=0$) we have a system of independent spins $\vec\sigma_i$, and in each MC sweep, we replace all spins $\vec\sigma_i$ with new ones from the equilibrium probability distribution. Therefore, the first MC sweeps are enough to equilibrate the system at the beginning of the algorithm, even for a sufficiently large inverse temperature $\beta$.

\section{Numerical Results and Discussion}\label{S4}
In this section we compare the performances of the algorithms introduced in the previous section. As the benchmark, we take the problem of minimizing the energy function of the random regular XORSAT problem with parameters $(K=4,L=3)$. Let us start from comparing the success probability of the $1$-local QR algorithm with that of the standard QA algorithm. 

\begin{figure}
\includegraphics[width=16cm]{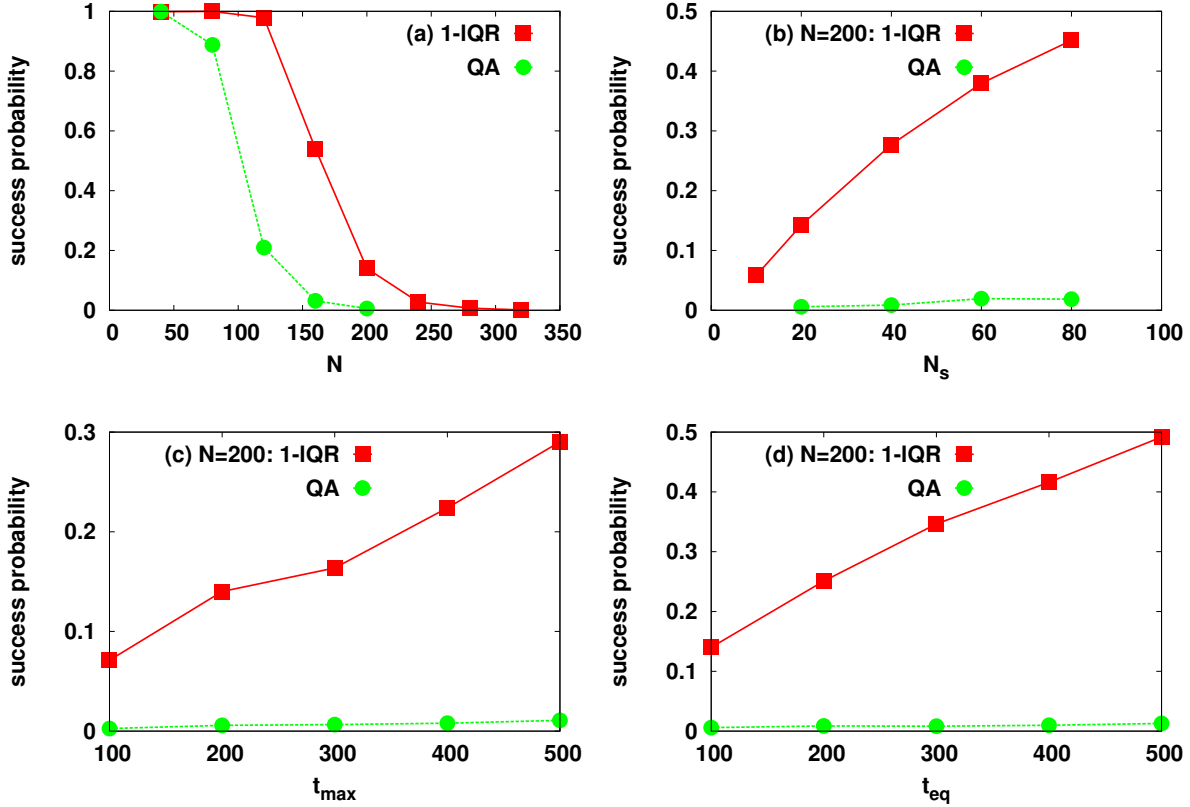} 
\caption{Comparing the success probabilities $P_{success}$ of the standard quantum annealing (QA) algorithm and the local quantum reinforcement ($1$-lQR) algorithm. The algorithm parameters, unless mentioned otherwise, are: $N_s=20, \beta=30, \Gamma=2, t_{max}=200, t_{eq}=100$ with $t_{av}=t_{eq}/2$ and $\delta r=0.002$ for the QR algorithm. $P_{success}$ vs (a) the number of variables $N$, (b) the number of imaginary-time slices $N_s$, (c) the maximum number of time steps $t_{max}$, and (d) the equilibration time $t_{eq}$. The data are obtained from (depending on the problem size) $2000$ to $10000$ runs of the algorithm on independent realizations of the problem.} \label{f1}
\end{figure}

Figure \ref{f1} shows the success probability of the two algorithms for different relevant parameter values in the algorithms. The success probability $P_{success}$ here refers to the fraction of times that an algorithm provides a zero-energy spin configuration satisfying all the constraints. Each time we take an independently generated random instance of the problem, which is identified with the random structure of the interaction graph and the random values of the couplings $J_a$. We run the algorithms for a sufficiently large number of problem instances $N_{samples}$ to obtain a reasonable stationary value for the success probability. The number of samples ranges from a few hundreds to at most ten thousands depending on the problem size,  As expected, we observe that $P_{success}$ decreases exponentially with the problem size $N$. The QR algorithm, however, exhibits much better performances than the QA algorithm for different parameter values. We recall that by adding the reinforcement to the Hamiltonian we are in fact increasing the minimal energy gap of the system in the annealing process \cite{QR-pra-2017}; that is because the reinforced Hamiltonian is assigning lower energies to the more probable states.

\begin{figure}
\includegraphics[width=16cm]{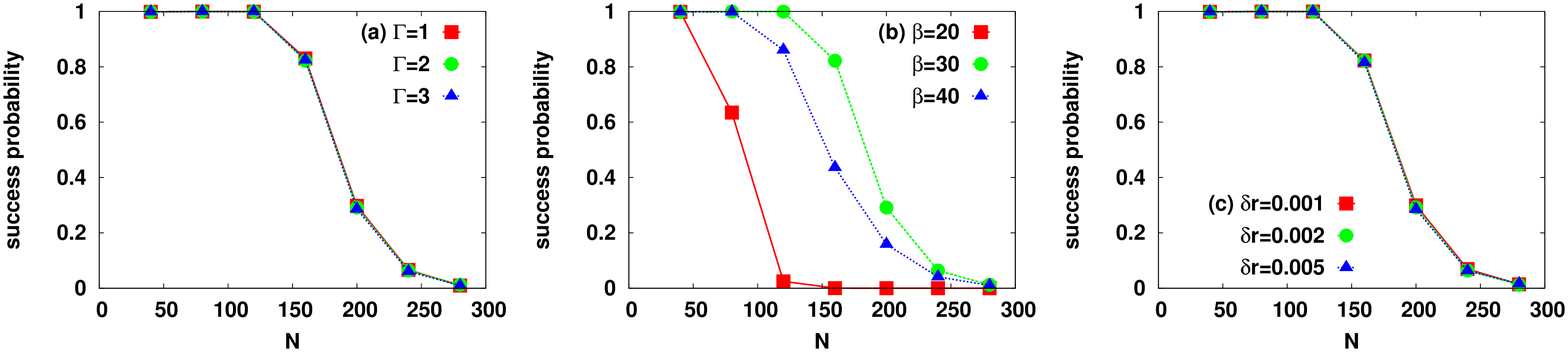} 
\caption{Success probabilities of the one-local QR algorithm. $P_{success}$ vs the problem size $N$ for different (a) transverse fields $\Gamma$, (b) inverse temperatures $\beta$, and (c) rates of increasing the reinforcement parameter. The algorithm parameters are $N_s=20, t_{max}=100, t_{eq}=500, t_{av}=400, \Gamma=2, \beta=30$, and $\delta r=0.002$, unless it is explicitly mentioned.} \label{f2}
\end{figure}

\begin{figure}
\includegraphics[width=16cm]{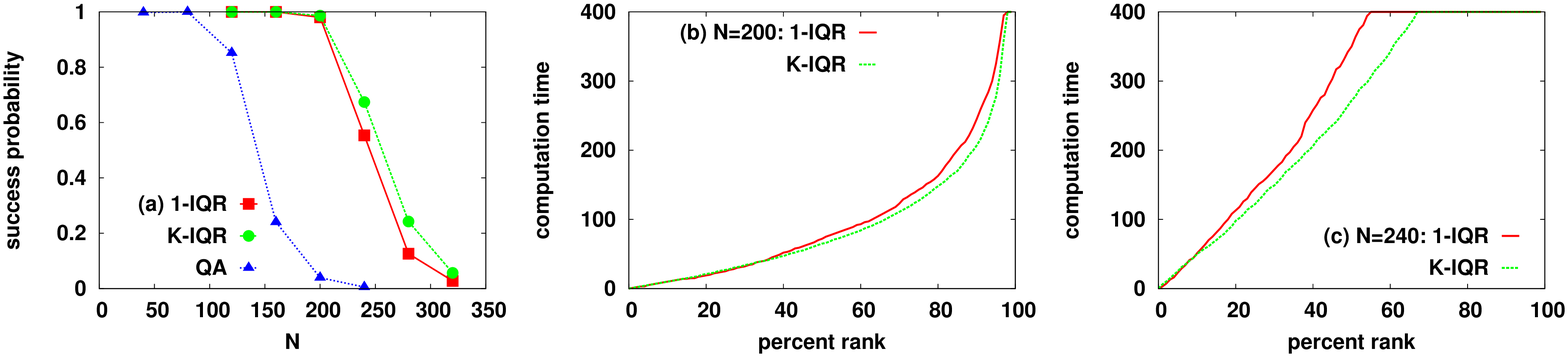} 
\caption{The algorithms performances for larger real and imaginary times. Here $t_{max}=400$, $N_s=60$, $\beta=90$, $\Gamma=2$, and $t_{eq}=500$. The success probability (a) and the percentile value of the computation time (number of time steps) in the $1$-lQR and $K$-lQR algorithms for $N=200$ (b), $N=240$ (c). In the QR algorithms $\delta r=0.001$ and $t_{av}=400$. The computation times are obtained from $2000$ independent problem instances.} \label{f3}
\end{figure}

\begin{figure}
\includegraphics[width=16cm]{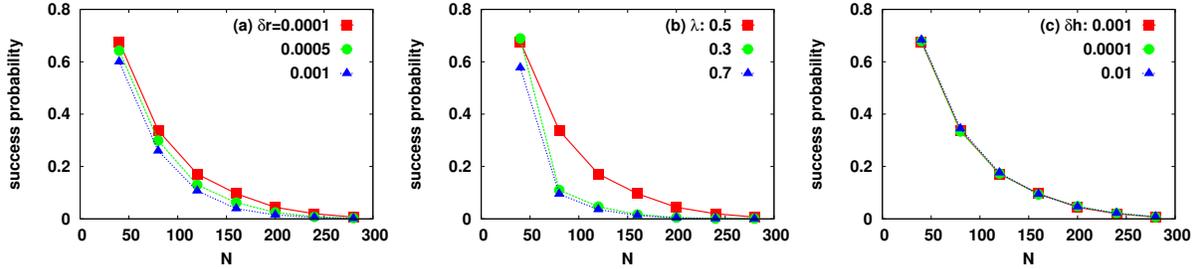} 
\caption{Success probability of the classical reinforced BP algorithm. $P_{success}$ for different (a) rates of increasing the reinforcement parameter $\delta r$, (b) damping parameters $\lambda$ in updating the messages, and (c) noise levels $\delta h$ in the external fields to break the problem symmetry. The data are obtained from $10^4$ runs of the algorithm on independent realizations of the problem. The maximum number of iterations is $t_{max}=10^5$. The other parameters (if not fixed) are $\delta r=0.0001, \lambda=0.5, \delta h=0.001$.} \label{f4}
\end{figure}

Figure \ref{f2} displays more results from the $1$-local QR algorithm to see how the algorithm parameters affect the success probability. Note that for $N_s=20$ the best performances are observed for $\beta=30$ (i.e., $\tau=1.5$). Moreover, the behavior of the algorithm is not very sensitive to the values of $\Gamma=1,2,3$ and $\delta r=0.001,0.002,0.005$. In Fig. \ref{f3}, we compare the efficiencies of the $1$-local and $K$-local QR algorithms for a larger number of imaginary-time slices and longer annealing times. We observe a small improvement in the success probability and computation time of the local QR algorithm by considering the $K$-local interactions in the wave function. Here, the quality of the approximate inverse algorithm in the $K$-local algorithm is very crucial. The difference in the performances of the two local algorithms is expected to be more pronounced if we employ more accurate inverse algorithms. Finally, for comparison, in Fig. \ref{f4} we also report the success probability of a powerful classical optimization algorithm (reinforced BP), which is described in the Appendix. This shows that by adding a local reinforcement to the quantum annealing algorithm, one can achieve performances that are better than or comparable to those of the classical algorithm.

\section{Conclusion}\label{S5}
We employed the path-integral quantum Monte Carlo to simulate the behavior of the quantum reinforcement algorithms in optimization of a hard constraint satisfaction problem. We observed that local quantum reinforcements can significantly improve the success probability of the standard quantum annealing algorithm. The performance of the simulated quantum reinforcement algorithm can systematically be improved by considering more accurate representations of the system wave function (e.g., tensor networks \cite{tn-siam-2008,tn-anp-2014,tn-arxiv-2018}) in the annealing process, and by utilizing more efficient approximations for estimating the wave-function parameters from the measurements.   
 
In this paper, we assumed the quantum system is close to the thermal equilibrium as the Hamiltonian changes with time. This means that in practice the equilibration time should be smaller than the time scale of changing the Hamiltonian. Moreover, we did not consider the effect of measurements, which are needed for implementing the reinforcement, on the quantum state of the system. In this sense, the simulated quantum reinforcement algorithm which was presented in this paper is closer to a classical optimization algorithm. A more realistic simulation of the quantum annealing process with reinforcement, should consider an open quantum system which also interacts with a classical (or even quantum) controller. The controller is to adjust the reinforcement Hamiltonian, which  depends on the outcomes of the necessary (weak) measurements, e.g., measurements of the local magnetizations \cite{DK-pra-1999,Qestimation-prl-2006,Qcontrol-book}. This is the subject of our future study. 

There are quantum annealers that provide hardware support for solving an optimization spin-glass problem \cite{qa-nature-2011,qa-nc-2013}. An experimental implementation of the quantum annealing algorithm on such devices first needs a mapping of the optimization problem to the Ising model with two-spin interactions \cite{lucas-fn-2014}. In addition, one also needs to embed the interactions of the Ising Hamiltonian onto the interaction graph of the specified device. Each of the above steps requires a polynomial number of auxiliary spins to be added to the system, and thus increases the size of the necessary device \cite{choi-qi-2008,choi-qi-2011,sg-prx-2015}. The quantum reinforcement algorithm could increase this complexity by adding other ancillary spins to the system for an indirect or weak measurement of the local magnetizations and correlations.


\appendix

\section{Bethe approximation and Belief Propagation (BP) equations}\label{app-1}
Consider an interacting system of $N$ binary variables $\sigma_i\in \{-1,+1\}$ with the following energy function
\begin{align}
E(\boldsymbol\sigma)=-\sum_{i=1}^N h_i\sigma_i-\sum_{a=1}^M h_a \prod_{i\in \partial a}\sigma_i.
\end{align}
The interaction pattern of the variables is identified with the neighborhood subsets $\partial a$ and $\partial i$. Here, $\partial a$ gives the set of variables in constraint $a$, and $\partial i$ is the set of constraints involving variable $i$.  The partition function for this problem reads as follows,
\begin{align}
Z=\sum_{\boldsymbol\sigma}e^{-E(\boldsymbol\sigma)},
\end{align}
where the inverse temperature parameter is absorbed in the couplings $h_{i,a}$. 
 
Assuming that the interaction graph is locally treelike, the local averages $m_i=\langle \sigma_i\rangle$ and $m_a=\langle \prod_{i\in \partial a}\sigma_i\rangle$ can be written in terms of the cavity probabilities \cite{MM-book-2009}, 
\begin{align}
m_i &=\frac{1}{z_i}\sum_{\sigma_i}\sigma_i e^{h_i\sigma_i} \prod_{a\in \partial i}\mu_{a\to i}(\sigma_i),\\
m_a &=\frac{1}{z_a}\sum_{\sigma_{\partial a}} e^{h_a\prod_{i\in \partial a}\sigma_i}\prod_{i\in \partial a}\left(\sigma_i\mu_{i\to a}(\sigma_i)\right).
\end{align}
Here $\mu_{i\to a}(\sigma_i)$ is the probability of state $\sigma_i$ for variable $i$ in the absence of interaction $a$, and $\mu_{a\to i}(\sigma_i)$ is the message that variable $i$ receives from interaction factor $a$ to satisfy the interaction. The $z_i$ and $z_a$ are normalization constants. The cavity messages are governed by the Bethe equations,
\begin{align}
\mu_{i\to a}(\sigma_i) &=\frac{1}{z_{i\to a}}e^{h_i\sigma_i} \prod_{b\in \partial i\setminus a}\mu_{b\to i}(\sigma_i),\\
\mu_{a\to i}(\sigma_i) &=\frac{1}{z_{a\to i}}\sum_{\sigma_{\partial a\setminus i}} e^{h_a\prod_{j\in \partial a}\sigma_j}\prod_{j\in \partial a\setminus i}\mu_{j\to a}(\sigma_j),
\end{align}
with the normalization constants $z_{i\to a}$ and $z_{a\to i}$. The cavity equations are solved by iteration starting from random initial values for the cavity messages. Then, the messages are used to find the local estimation values from the above equations.

\subsection{Solving the inverse problem within the Bethe approximation}\label{app-11}
The BP algorithm provides an efficient way of estimating the expectation values, given the energy function. This approximation method is useful also in solving the inverse problem of constructing the energy function, here the parameters $h_i$ and $h_a$, which best describes the given expectation values $m_i$ and $m_a$. A simple strategy, assuming that there is no error in the $m_{i,a}$, is to find the set of parameters $\mathbf{h}=\{h_{i,a}\}$ that minimize the differences between the resulting $m_{i,a}[\mathbf{h}]$ and the given values $m_{i,a}$. The following algorithm tries to solve the above problem with iteration:  

\begin{itemize}

\item Start at time step zero $t=0$ with initial parameters $h_i(t)=h_a(t)=0$. 

\item For $t=0,\dots,t_{inv}$ do:    

\end{itemize}
\begin{enumerate}

\item compute the expectation values $m_{i,a}[\mathbf{h}(t)]$;  

\item compute the deviations $\Delta_{i,a}=|m_{i,a}-m_{i,a}[\mathbf{h}(t)]|$;

\item stop if the maximum deviation is smaller than $\epsilon$;

\item change the parameters $h_{i,a}(t+1)=h_{i,a}(t)+\eta(m_{i,a}-m_{i,a}[\mathbf{h}(t)])$.  

\end{enumerate}

Here, we use the BP algorithm to estimate the average values $m_{i,a}[\mathbf{h}(t)]$. The parameter $\eta$ is a sufficiently small and positive number.

\subsection{The reinforced BP algorithm}\label{app-12}
The Bethe approximation also provides an approximate algorithm to find a solution to the XORSAT problem. A solution is a spin configuration which satisfies all the XORSAT constraints, i.e.  $\prod_{i\in \partial a}\sigma_i=J_a$ for all the $a$, with $J_a=\pm 1$. To this end, one introduces the reinforced term $E_r(\boldsymbol\sigma)=-r\sum_i \mu_i(\sigma_i)$ to the energy function.
The reinforcement parameter $r$ is assumed to increase slowly with the number of algorithm iterations. More precisely, the reinforced BP (rBP) equations for the cavity messages at iteration $t$ are
\begin{align}
\mu_{i\to a}^{t+1}(\sigma_i) &=\frac{1}{z_{i\to a}}e^{h_i\sigma_i+r(t)\mu_i^t(\sigma_i)} \prod_{b\in \partial i\setminus a}\mu_{b\to i}^t(\sigma_i),\\
\mu_{a\to i}^{t+1}(\sigma_i) &=\frac{1}{z_{a\to i}}\sum_{\sigma_{\partial a\setminus i}} \mathbb{I}_a(\sigma_{\partial a})\prod_{j\in \partial a\setminus i}\mu_{j\to a}^t(\sigma_j).
\end{align}
The small external fields $h_i$, with a magnitude much less than one, are to break the high symmetry of the problem. The indicator function $\mathbb{I}_a(\sigma_{\partial a})$ is one if constraint $a$ is satisfied, otherwise, it is zero. Moreover, the local marginal probabilities are given by 
\begin{align}
\mu_{i}^{t+1}(\sigma_i) =\frac{1}{z_i}e^{h_i\sigma_i+r(t)\mu_i^t(\sigma_i)} \prod_{a\in \partial i}\mu_{a\to i}^t(\sigma_i).
\end{align}

The equations are solved by iteration starting from random initial messages and updating them according to the above equations for at most $t_{max}$ iterations. At each iteration, we update all the cavity and local marginals. We also introduce damping to the iterative process, i.e., at each step the messages are updated as follows: $\mu^{t+1}=\lambda \mu^t+(1-\lambda)\mu^{t+1}$ with a damping parameter $0<\lambda<1$. We set $r(0)=0$ and increase the reinforcement parameter linearly with the number of iterations as $r(t+1)=r(t)+\delta r$. After each iteration, a candidate spin configuration for solution is obtained by looking at the local marginal probabilities $\sigma_i^*=\arg \max \mu_i(\sigma_i)$. The algorithm stops when the candidate configuration is a solution to the problem.


\begin{thebibliography}{prsty}

\bibitem{sat-nature-1999} Monasson, Remi, et al. "Determining computational complexity from characteristic ‘phase transitions’." Nature 400.6740 (1999): 133.

\bibitem{sat-science-2002} Mezard, Marc, Giorgio Parisi, and Riccardo Zecchina. "Analytic and algorithmic solution of random satisfiability problems." Science 297.5582 (2002): 812-815.


\bibitem{SB-book-1998} Sutton, Richard S., and Andrew G. Barto. Introduction to reinforcement learning. Vol. 135. Cambridge: MIT Press, 1998.

\bibitem{BZ-prl-2006} Braunstein, Alfredo, and Riccardo Zecchina. "Learning by message passing in networks of discrete synapses." Physical review letters 96.3 (2006): 030201.


\bibitem{gibbs-pnas-2007} Krzakala, Florent, et al. "Gibbs states and the set of solutions of random constraint satisfaction problems." Proceedings of the National Academy of Sciences 104.25 (2007): 10318-10323.

\bibitem{semerjian-jstat-2008} Semerjian, Guilhem. "On the freezing of variables in random constraint satisfaction problems." Journal of Statistical Physics 130.2 (2008): 251-293.

\bibitem{gibbs-pre-2008} Dall’Asta, Luca, Abolfazl Ramezanpour, and Riccardo Zecchina. "Entropy landscape and non-Gibbs solutions in constraint satisfaction problems." Physical Review E 77.3 (2008): 031118.



\bibitem{clustering-prl-2005} Mezard, Marc, Thierry Mora, and Riccardo Zecchina. "Clustering of solutions in the random satisfiability problem." Physical Review Letters 94.19 (2005): 197205.

\bibitem{clustering-jstat-2008} Montanari, Andrea, Federico Ricci-Tersenghi, and Guilhem Semerjian. "Clusters of solutions and replica symmetry breaking in random k-satisfiability." Journal of Statistical Mechanics: Theory and Experiment 2008.04 (2008): P04004.




\bibitem{xor-prl-2001} Franz, Silvio, et al. "Exact solutions for diluted spin glasses and optimization problems." Physical Review Letters 87.12 (2001): 127209.

\bibitem{xor-pre-2001} Ricci-Tersenghi, Federico, Martin Weigt, and Riccardo Zecchina. "Simplest random k-satisfiability problem." Physical Review E 63.2 (2001): 026702.


\bibitem{qxor-prl-2009} Harrow, Aram W., Avinatan Hassidim, and Seth Lloyd. "Quantum algorithm for linear systems of equations." Physical review letters 103.15 (2009): 150502.


\bibitem{MM-book-2009} Mezard, Marc, and Andrea Montanari. Information, physics, and computation. Oxford University Press, 2009.



\bibitem{entropy-jstat-2016} Baldassi, Carlo, et al. "Local entropy as a measure for sampling solutions in constraint satisfaction problems." Journal of Statistical Mechanics: Theory and Experiment 2016.2 (2016): 023301.

\bibitem{globalgame-2011} Ramezanpour, Abolfazl, John Realpe-Gomez, and Riccardo Zecchina. "Statistical physics approach to graphical games: local and global interactions." The European Physical Journal B 81.3 (2011): 327-339.

\bibitem{sign-prb-2012} Ramezanpour, A., and R. Zecchina. "Sign problem in the Bethe approximation." Physical Review B 86.15 (2012): 155147.



\bibitem{QR-pra-2017} Ramezanpour, A. "Optimization by a quantum reinforcement algorithm." Physical Review A 96.5 (2017): 052307.



\bibitem{ALZ-pra-1993} Aharonov, Yakir, Luiz Davidovich, and Nicim Zagury. "Quantum random walks." Physical Review A 48.2 (1993): 1687.

\bibitem{K-cp-2003} Kempe, Julia. "Quantum random walks: an introductory overview." Contemporary Physics 44.4 (2003): 307-327.

\bibitem{A-jqi-2003} Ambainis, Andris. "Quantum walks and their algorithmic applications." International Journal of Quantum Information 1.04 (2003): 507-518.




\bibitem{F-sci-2001} Farhi, Edward, et al. "A quantum adiabatic evolution algorithm applied to random instances of an NP-complete problem." Science 292.5516 (2001): 472-475.

\bibitem{NC-book-2002} Nielsen, Michael A., and Isaac Chuang. "Quantum computation and quantum information." (2002).




\bibitem{lloyd-prl-1998} Abrams, Daniel S., and Seth Lloyd. "Nonlinear quantum mechanics implies polynomial-time solution for NP-complete and $\#$ P problems." Physical Review Letters 81.18 (1998): 3992.



\bibitem{AHJ-pnas-2010} Altshuler, Boris, Hari Krovi, and Jeremie Roland. "Anderson localization makes adiabatic quantum optimization fail." Proceedings of the National Academy of Sciences 107, no. 28 (2010): 12446-12450.

\bibitem{qxor-prl-2010} Jorg, Thomas, et al. "First-order transitions and the performance of quantum algorithms in random optimization problems." Physical review letters 104.20 (2010): 207206.

\bibitem{qxor-pre-2011} Hen, Itay, and A. P. Young. "Exponential complexity of the quantum adiabatic algorithm for certain satisfiability problems." Physical Review E 84.6 (2011): 061152.

\bibitem{qxor-pra-2012} Farhi, Edward, et al. "Performance of the quantum adiabatic algorithm on random instances of two optimization problems on regular hypergraphs." Physical Review A 86.5 (2012): 052334.




\bibitem{B-jpa-2009} Berry, M. V. "Transitionless quantum driving." Journal of Physics A: Mathematical and Theoretical 42.36 (2009): 365303.

\bibitem{C-prl-2013} del Campo, Adolfo. "Shortcuts to adiabaticity by counterdiabatic driving." Physical review letters 111.10 (2013): 100502.


\bibitem{localCA-pra-2014} Saberi, Hamed, et al. "Adiabatic tracking of quantum many-body dynamics." Physical Review A 90.6 (2014): 060301.


\bibitem{localCA-pnas-2017} Sels, Dries, and Anatoli Polkovnikov. "Minimizing irreversible losses in quantum systems by local counterdiabatic driving." Proceedings of the National Academy of Sciences (2017): 201619826.



\bibitem{tosatti-prb-2002} Martonak, Roman, Giuseppe E. Santoro, and Erio Tosatti. "Quantum annealing by the path-integral Monte Carlo method: The two-dimensional random Ising model." Physical Review B 66.9 (2002): 094203.

\bibitem{pathMC-prb-2008} Krzakala, Florent, et al. "Path-integral representation for quantum spin models: Application to the quantum cavity method and Monte Carlo simulations." Physical Review B 78.13 (2008): 134428.

\bibitem{QC-prep-2013} Bapst, Victor, et al. "The quantum adiabatic algorithm applied to random optimization problems: The quantum spin glass perspective." Physics Reports 523.3 (2013): 127-205.




\bibitem{inverse-advanc-2017} Nguyen, H. Chau, Riccardo Zecchina, and Johannes Berg. "Inverse statistical problems: from the inverse Ising problem to data science." Advances in Physics 66.3 (2017): 197-261.













\bibitem{tn-siam-2008} Markov, Igor L., and Yaoyun Shi. "Simulating quantum computation by contracting tensor networks." SIAM Journal on Computing 38.3 (2008): 963-981.

\bibitem{tn-anp-2014} Orus, Roman. "A practical introduction to tensor networks: Matrix product states and projected entangled pair states." Annals of Physics 349 (2014): 117-158.

\bibitem{tn-arxiv-2018} Kourtis, Stefanos, et al. "Fast counting with tensor networks." arXiv preprint arXiv:1805.00475 (2018).



\bibitem{DK-pra-1999} Doherty, Andrew C., and Kurt Jacobs. "Feedback control of quantum systems using continuous state estimation." Physical Review A 60.4 (1999): 2700.

\bibitem{Qestimation-prl-2006} Smith, Greg A., et al. "Efficient quantum-state estimation by continuous weak measurement and dynamical control." Physical review letters 97.18 (2006): 180403.


\bibitem{Qcontrol-book} Wiseman, Howard M., and Gerard J. Milburn. Quantum measurement and control. Cambridge university press, 2009.




\bibitem{qa-nature-2011} Johnson, Mark W., et al. "Quantum annealing with manufactured spins." Nature 473.7346 (2011): 194.

\bibitem{qa-nc-2013} Boixo, Sergio, et al. "Experimental signature of programmable quantum annealing." Nature communications 4 (2013): 2067.



\bibitem{lucas-fn-2014} Lucas, Andrew. "Ising formulations of many NP problems." Frontiers in Physics 2 (2014): 5.


\bibitem{choi-qi-2008} Choi, Vicky. "Minor-embedding in adiabatic quantum computation: I. The parameter setting problem." Quantum Information Processing 7.5 (2008): 193-209.

\bibitem{choi-qi-2011} Choi, Vicky. "Minor-embedding in adiabatic quantum computation: II. Minor-universal graph design." Quantum Information Processing 10.3 (2011): 343-353.

\bibitem{sg-prx-2015} Venturelli, Davide, et al. "Quantum optimization of fully connected spin glasses." Physical Review X 5.3 (2015): 031040.



\end{thebibliography}
\end{document}